\documentclass{eas}
\usepackage{graphicx}
\begin{document}

\TitreGlobal{Mass Profiles and Shapes of Cosmological Structures}

\title{Velocity moments of dark matter haloes}
\author{Wojtak, R.}\address{Astronomical Observatory of Jagiellonian University, Cracow, Poland}
\author{{\L}okas, E.}\address{Nicolaus Copernicus Astronomical Center, Warsaw, Poland}
\author{Gottl\"ober, S.}\address{Astrophysikalisches Institut Potsdam, Germany}
\author{Mamon, G.}\address{Institut d'Astrophysique de Paris, France}
\runningtitle{Velocity moments of dark matter haloes}
\setcounter{page}{1}
\index{Wojtak, R.}
\index{{\L}okas, E.}
\index{Gotl\"ober, S.}
\index{Mamon, G.}

\maketitle
\begin{abstract}
Using cosmological N-body simulations we study the
line-of-sight velocity distribution of dark matter
haloes focusing on the lowest-order even moments,
dispersion and kurtosis, and their application to
estimate the mass profiles of cosmological structures.
For each of the ten massive haloes selected from the simulation box
we determine the virial mass, concentration and the anisotropy parameter.
In order to emulate observations from each halo we choose randomly 300 particles
and project their velocities and positions along the line of sight and on the surface
of the sky, respectively. After removing interlopers we calculate the profiles of the line-of-sight
velocity moments and fit them with the solutions of the Jeans equations.
The estimates of virial mass, concentration parameter and velocity anisotropy
obtained in this way are in good agreement with the values found from the full 3D analysis.
\end{abstract}
%
The purpose of this study was to verify how well the global properties of dark matter
haloes can be reproduced from the joint fitting of the line-of-sight velocity dispersion and kurtosis
profiles to the solutions of the Jeans equations, a method first applied by {\L}okas \& Mamon (2003)
to the Coma cluster. Our approach is similar to that of Sanchis et al. (2004).
We have used ten massive haloes (with masses larger than $10^{14}$ M$_{\odot}$) extracted from
a cosmological simulation of the standard $\Lambda$CDM model performed in a box of size
$150$ Mpc$/h$ (Wojtak et al. 2005). We determined the properties of the haloes estimating
their virial mass, concentration (as defined in the NFW formula) and anisotropy parameter.

\begin{figure}[t]
   \centering
   \includegraphics[width=12.75cm]{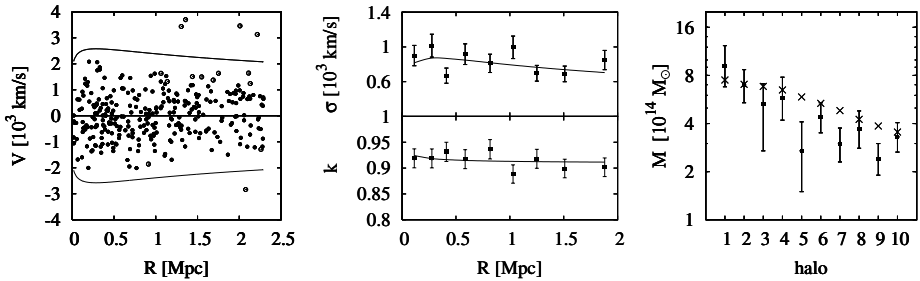}
      \caption{Left panel: the velocity diagram with filled (empty)
      circles denoting bound (unbound) particles and curves
      separating interlopers from members. Middle panel: velocity
      dispersion and kurtosis variable $k$ with curves showing the fitted
      solutions of the Jeans equations. Right panel: virial mass estimates
      for ten haloes obtained in 3D analysis (crosses) and in the Jeans analysis
      (squares).}
       \label{figure1}
\end{figure}

To mimic observations
we project velocities and positions of the halo particles along the line of sight
of an imaginary observer placed at the distance of 100 Mpc from a halo and onto the plane
of the sky respectively, producing velocity diagrams $V(R)$. For each halo we choose
randomly 300 particles with velocities in the range $+/-$ 4000 km/s with respect to
the halo mean velocity.
In order to remove interlopers from our samples for each halo we compute
the line-of-sight velocity dispersion profile $\sigma(R)$ and fit it with
solutions of the Jeans equation assuming isotropic orbits.
We reject as interlopers the particles with velocities outside the fitted
range  $+/-3\sigma(R)$
with respect to the halo mean velocity, repeating the procedure until convergence is
achieved. An example of the velocity diagram for the most massive halo in our sample
is shown in the left panel of Fig.\ref{figure1} together with the results of this
procedure.

After removing interlopers we measure final profiles of the line-of-sight velocity moments:
dispersion and kurtosis. Then we fit these data with solutions of the Jeans
equations ({\L}okas \& Mamon 2003) obtaining estimates of virial mass,
concentration and anisotropy parameter. The middle panel of
Fig.\ref{figure1} shows the velocity dispersion and kurtosis variable $k$
(related to the estimator of kurtosis $K$ by $k=[\log (3K/2.68) ]^{1/10}$) for our most massive
halo.

The fitted parameters are in good agreement with values obtained from the full
3D information. As an example we show in the right panel of Fig.\ref{figure1}
values of the virial mass for the whole sample of ten haloes found in the full 3D analysis
(crosses) and in the Jeans analysis (squares with errors following
from the sampling errors of velocity moments). We
conclude that the method of joint fitting of line-of-sight velocity dispersion
and kurtosis reproduces basic parameters of dark matter haloes
in a satisfactory way.

\end{document}